\documentclass[dvipsnames, 12pt]{article}
\usepackage{tabularx}
\usepackage{array}
\usepackage{siunitx}   
\usepackage[margin=25mm]{geometry}
\usepackage{array}
\usepackage{siunitx}
\usepackage{multirow}
\usepackage[sort,comma,authoryear]{natbib}
\usepackage{authblk}
\usepackage{caption}
\usepackage{framed}
\usepackage{xr-hyper}
\usepackage{hyperref}
\usepackage{lscape}
\usepackage{ragged2e}
\hypersetup{
    colorlinks=true,
    linkcolor=blue,
    filecolor=blue,      
    urlcolor=blue,
    citecolor=blue,
    breaklinks=true}
\usepackage{color,soul}
\usepackage{graphicx}
\usepackage{amsmath}
\usepackage{amssymb }
\usepackage{setspace}
\usepackage{threeparttable}
\usepackage{booktabs}
\usepackage{subcaption}
\usepackage{diagbox}
\usepackage{bm}
\usepackage{algorithm}
\usepackage{algpseudocode}

\onehalfspace

\title{Affective Polarization Amongst Swedish Politicians}

\author[a]{François t'Serstevens \thanks{Corresponding Author: François t'Serstevens. f.tserstevens@maastrichtuniversity.nl}} 
\author[b]{Roberto Cerina}
\author[c]{Gustav Peper}

\affil[a]{\footnotesize Department of Data Analytics and Digitalisation, Maastricht University, 53 Tongersetraat, 6211 LM Maastricht, Netherlands}
\affil[b]{Department of Media Studies, University of Amsterdam, 1090 GN Amsterdam Turfdraagsterpad 9}
\affil[c]{Independent Researcher - gustav.peper@gmail.com}

\date{\today}

\begin{document}
\maketitle
\begin{abstract}

This study investigates affective polarization among Swedish politicians on Twitter from 2021 to 2023, including the September 2022 parliamentary election. Analysing over 25,000 tweets and employing large language models (LLMs) for sentiment and political classification, we distinguish between positive partisanship (support for one’s own side) and negative partisanship (criticism of opponents). 

Our findings are contingent on the definition of the in-group. When political in-groups are defined at the ideological bloc level, negative and positive partisanship occur at similar rates. 
However, when the in-group is defined at the party level, negative partisanship becomes significantly more dominant and is $1.51$ times more likely $[1.45, 1.58]$. 
This effect is even stronger among extreme politicians, who engage in negativity more than their moderate counterparts.  
Negative partisanship also proves to be a strategic choice for online visibility, attracting $3.18$ more likes and $1.69$ more retweets on average. 

By adapting methods developed for two-party systems and leveraging LLMs for Swedish-language analysis, we provide novel insights into how multiparty politics shapes polarizing discourse. Our results underscore both the strategic appeal of negativity in digital spaces and the growing potential of LLMs for large-scale, non-English political research.

\end{abstract}
\newpage
\section{Introduction}

Affective polarization, characterized by intensifying hostility toward opposing political parties and stronger positive feelings toward one’s own, has become an increasingly prominent trend in Western democracies since the 1980s \citep{abramowitz2016rise}. 
The emergence of partisan media outlets and the fragmentation of the media landscape have fueled this polarization, amplifying ideological differences and undermining the spirit of compromise \citep{benkler2018network, abramowitz2019united}. Scholars and commentators widely view the rise of affective polarization as a significant threat to democratic governance, particularly in its manifestation as negative partisanship—characterized by pervasive negative attitudes toward opposing parties. By prioritizing the defeat of political adversaries over constructive policymaking, negative partisanship erodes incentives for collaboration and fosters an adversarial political climate in which partisan victories take precedence over addressing critical societal challenges \citep{brownstein2021trump, TheEconomist2024, abramowitz2016rise}.

This phenomenon, known as affective polarization, has been extensively studied in the U.S., where political divisions are sharply defined \citep{yu2024partisanship, lee2022negative}. On a smaller scale, scholars have extended these investigations to the European Union (EU), finding evidence of it there as well. However, because EU countries have fundamentally different political systems than the US, the intensity of affective polarization is not constant across countries \citep{reiljan2020fear, torcal2022affective}. 
The multi-party structure of most EU member states also requires different methodological approaches, as tools like DW-NOMINATE and VoteView \citep{lewis2019voteview}, which provide estimations of political ideologies in two-party systems, struggle to capture the complexity of political divides in such settings.

Given these challenges, this paper aims to replicate the work \cite{yu2024partisanship} in the context of the EU. 
Their study analyses political partisanship on social media, focusing on the engagement of American politicians and citizens with partisan content. It finds that, in general, politicians tend to engage more in positive partisanship -- supporting their own party, rather than negative partisanship -- attacking the opposition. The exception to this trend is ideologically extreme politicians, who are more inclined to engage in negative partisanship.
Building on their framework, we investigate how negative partisanship manifests on Twitter among prominent figures in Swedish politics. 
Sweden provides an interesting case due to its proportional representation system and multi-party landscape, which differs significantly from the U.S. two-party system. 
Specifically, we examine how negative and positive partisanships manifest in a multi-party context and whether these strategies are effective for Swedish political elites.

We also contribute to the growing literature on leveraging large language models (LLMs) in non-Anglophone settings, particularly for studying political polarization \citep{neplenbroek2024mbbq, barbereau2024gpt}. While LLMs offer powerful tools for political analysis \citep{tornberg2024large}, they are not neutral observers; research has shown that their political classifications often reflect biases inherent in their training data or reinforcement learning processes \citep{hartmann2023political, abid2021persistent}. Given these challenges, we focus on setting realistic expectations for LLM-based political classification in Sweden.

This paper is structured as follows: After briefly explaining the relevant concepts of this article, we start by describing the data collection process, where we use the Twitter API to gather tweets from prominent Swedish political figures in Section \ref{sc:method}. 
Next, we outline the methodology for assessing the sentiment and partisanship of $\mathbb{X}$ users and their tweets.
We then explain the analytical tools used and their results in Section \ref{sc:results}. Section \ref{sc:discussion} discusses the implications of our results for understanding political polarization on social media in Sweden’s multi-party system.

\subsection{Theoretical Framework}

Affective polarization refers to individuals' emotions towards members of their own political party or group becoming more positive while their feelings towards members of the opposing party or group become more negative \citep{iyengar2019origins}. It consists of two main components, positive partisanship (PP) and negative partisanship (NP) and has progressively become more common over the last decades \citep{falkenberg2022growing, torcal2022affective,kawecki2022end}.
Positive partisanship describes political attitudes or behaviours driven primarily by strong support for one’s own political party or ideological group. It emphasizes in-group affinity, characterized by loyalty and attachment to one’s party, without necessarily involving hostility toward opposing groups  \citep{brewer1999psychology}. 
In contrast, negative partisanship describes out-group disdain, where individuals develop negative attitudes or hostility toward rival parties or ideological groups. 
Affective polarisation has been predominately studied as a unified concept \citep{reiljan2020fear, torcal2022affective}, but \cite{anderson2022us} suggest that negative and positive partisanship are not exclusive to one another.

The effects of negative and positive partisanships on voting behaviour and, more generally, political stance has been a core discussion amongst scholars. In the United States, \cite{lee2022negative} and \cite{yu2024partisanship} find that positive partisanship is the dominant form of partisanship, whilst \cite{iyengar2018strengthening} suggest that the party animosity eclipses positive affect as a motive for political participation, both research groups using the American National Election Studies (ANES) dataset.

In the EU, affective polarisation and particularly negative partisanship are present too \citep{hahm2023divided}. \cite{reiljan2020fear} argues that the degree of affective polarisation is heterogeneous across political systems. Specifically, he argues that Northwestern European countries are more moderate in terms of partisan feelings than their Southern and Central Eastern countries, some political systems being more prone to interparty hostility. 
In practice, \cite{knudsen2021affective} finds that affective polarisation levels in the northern EU countries are similar to the US's despite the multi-party settings. Affective polarisation has a definite effect on voting behaviour, though it is heterogeneous across political parties and countries \citep{mayer2017negative}.  One challenge in studying affective polarization is the lack of widely validated measures, many of which have primarily been developed and tested in the U.S. context \citep{wagner2024affective, gidron2022validating}.

Part of the difficulty in studying affective polarisation stems from the difficulty of measuring it accurately. Self-reported metrics, while commonly used, come with significant limitations \cite{guess2019accurate}. Similarly, experimental studies, may face challenges when extending their findings beyond controlled settings \citep{morton2008experimentation, druckman2006growth}. As \cite{yu2024partisanship} we address these limitations by analysing posts from elite politicians on $\mathbb{X}$. We believe that social media data is well suited to this research because it offers a direct communication channel to the electorate. In the last decades it has profoundly influenced democratic processes \cite{tucker2017liberation}, with elite communication dominating these platforms \cite{barbera2019leads}. Despite its merits, this approach is not without limitations. Polarizing rhetoric from elites on social media does not necessarily imply equivalence among the public \citep{lee2022negative}. Thus, the results presented here reflect the state of elite affective polarization, not public sentiment.

Political elites on social media differ significantly from laypeople in their experiences and reactions. They are more likely to encounter attacks or incivility and, in turn, respond in a hostile manner \citep{munger2021don, iyengar2018media}. For instance, approximately 18\% of all tweets mentioning legislators are uncivil \citep{theocharis2020dynamics}. This trend is further heightened by increasing ideological polarization among political elites, which suggests that negative partisanship is becoming a dominant response in these interactions \citep{theriault2008party, webster2017ideological}. 
The prevalence of incivility among elites can partially be attributed to the strategic advantages it confers. By criticizing their opponents, politicians can bolster their own standing among their base \citep{herbst2010rude, ballard2022incivility}. However, one might argue that coalition politics, which often necessitate cross-party collaboration, could discourage overt hostility. The Swedish political system, characterized by coalition governance, provides an interesting case in point. Despite this structure, \citet{reiljan2020fear} suggests that affective polarization can be just as strong, if not stronger, in multi-party systems compared to the United States. In the Nordic context, Sweden exhibits particularly high levels of affective polarization, while Norway's levels are more comparable to those in the U.S. \citep{ryan2023exploring,knudsen2021affective}.\\
Beyond the strategic dimensions, the increased visibility of negative posts may also drive elites toward negative partisanship. According to the negativity bias theory, humans are more likely to respond to negative stimuli than to positive ones \citep{rozin2001negativity, soroka2019cross}. This implies that negative partisanship is more likely to generate engagement on social media, such as retweets and likes, than its positive counterpart, as verified by \cite{rathje2021out} in the US. Based on these considerations, we formulate two hypotheses on negative partisanship in Sweden:

\textbf{H1:} Political elites are likelier to engage in negative rather than positive partisanship. \\
\textbf{H2:} Negative partisanship is more likely to gather reactions than its positive counterpart.

Beyond the prevalence of negative partisanship, we also anticipate that it is more common among extremist politicians as found by \cite{yu2024partisanship} in the US.
From a theoretical perspective, the negative feedback loop between political polarization and negative partisanship suggests that extreme partisans are more likely to engage in negative partisanship \citep{ost2004politics, su2018uncivil}. Moreover, extremists often feel that their views are underrepresented in mainstream media \citep{hong2019politicians}. As a result, extreme political elites may turn to political incivility, as it helps their message stand out and resonate more strongly with their core supporters, who may feel marginalized by conventional political discourse.
From a practical standpoint \cite{ballard2022incivility} suggest that politicians with extreme viewpoints, along with those in the opposition and those with safe political positions are more likely to engage negative partisanship. Likewise, \cite{widmann2021emotional} ties the usage negative emotional language to populist parties and joyful language to more moderate parties. Based on these insights, we hypothesize that extreme political elites are more likely to engage in negative partisanship compared to their moderate counterparts.

\textbf{H3:} Extreme political elites are more likely to use negative partisanship than their moderate counterparts.

\section{Methods and Materials} \label{sc:method}

This section details the data collection process and analytical tools used in the study. We collected tweet data from the X API and analysed sentiment using multiple (LLMs). Politician data was compiled from multiple sources, including parliamentary records and expert surveys.
All data, along with the R, Python, and Stan code used for collection and analysis, are available on GitHub (https://github.com/ftserstevens/Sweden).

\subsection{Tweet Data}

\begin{figure}[]
    \centering
    \begin{minipage}{0.4\textwidth}
        \centering
        \includegraphics[width=\linewidth]{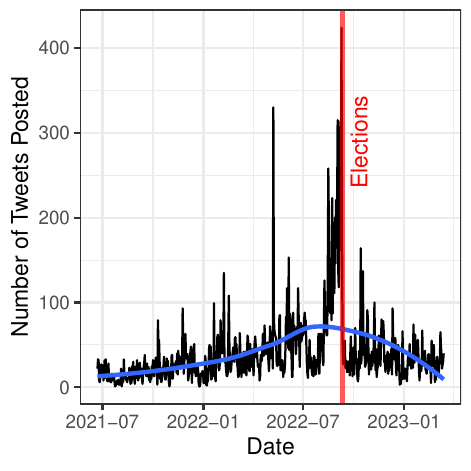}
        \caption{Amount of tweets posted by date.}
        \label{fig:TweetsbyDate}
    \end{minipage}

\end{figure}

We collected $25,129$ tweets posted by $79$ leading Swedish political figures between 22-06-2021 and 2023-03-15. Notably, some prominent Swedish political figures did not have private Twitter accounts during that period\footnote{The full list of selected politicians can be found in the online appendix.}, e.g. Ulf Kristersson, leader of the Moderate party (2017) and prime minister (2022). We chose this period as the Swedish parliamentary elections took place in September 2022, with a new ruling coalition being formed shortly thereafter. The inclusion of this regime change allows us to study in- and out-government animosity online as was done by \citet{yu2024partisanship}. Of the collected tweets, $44.45\%$ were posted by the left bloc and $55.55\%$ were posted by the right bloc. Figure \ref{fig:TweetsbyDate} depicts the tweeting rate over the time period of our sample. As expected, it displays more activity close to the national elections of September 2022, with large peaks right before and after the elections.

\subsection{Tweet Sentiment}
\label{sc:TweetSent}

After downloading the necessary tweets, the next step is scoring their sentiment. To do so, we leverage 4 different LLM-based approaches. First, we estimate the sentiment with an open-source Swedish LLM. Second,  we used an English-based open-source LLM to score the sentiment, translating the tweets beforehand. Third, we use GPT-4o to rate the sentiment in English and Swedish. This multi-model approach allows us to study the degree to which non-English text can be successfully analysed using LLMs and to capture the sensitivity of sentiment analysis to variations in the amount of non-English training data. All approaches estimate three classes of sentiment - \textit{Positive}, \textit{Neutral} and \textit{Negative}. 

\begin{enumerate}
    \item \textbf{Native Swedish Scoring}. The first approach uses \texttt{KBLab/robust-swedish-sentim-\\ent-multiclass}, a multi-label transformer-based sentiment classifier trained on Swedish texts \citep{hägglöf2023a}. We selected this model because it was trained on 75K texts from multiple domains, among them 12K tweets. It is robust in the sense that it is trained on multiple datasets of different text types.
    \item \textbf{English Translation Scoring}. In this approach, we first translate Swedish tweets into English using \texttt{facebook/nllb-200-distilled-600M} and then score their sentiment using \texttt{cardiffnlp/twitter-roberta-base-sentiment-latest} \citep{camacho2022tweetnlp}, a model specifically fine-tuned for sentiment analysis on tweet data. This method is motivated by the availability of high-performance, fine-tuned models in English, which are tailored to the unique characteristics of Twitter data.
    \item \textbf{GPT Scoring}. In this last method, we score the sentiment of the tweets using OpenAI's \texttt{gpt-4o} model.  To do so we give the following prompt \textit{``As an AI with expertise in language and emotion analysis, your task is to analyze the sentiment of the following text. Return only the class of the sentiment either positive, neutral or negative''}, we translate the prompt into Swedish for the Swedish sentiment estimation.
\end{enumerate}

To evaluate the quality of the models the research team rated the sentiment $500$ tweets randomly selected from the initial pool of tweets in their original language. The models were evaluated based on their accuracy in predicting the correct sentiment class compared to the manually annotated ground truth. We present the F1 scores and Balanced Accuracies of all models because of the high imbalance in sentiment classes within the sample, i.e. there are many more neutrals and negatives than positives. It is worth noting that the metrics presented here are for a three-class classification task (Positive, Neutral and negative), unlike their more common use for two-class classification. 


The results, displayed in Table \ref{Tab: Sentiment} show that Swedish-native models perform better overall, with the GPT Swedish model achieving the highest F1 score (0.696) and Balanced Accuracy (0.696). This indicates that language-specific models, like the Swedish GPT, are better suited for sentiment analysis of Swedish tweets compared to translation-based or English models. The Open-Source Swedish model also outperforms the Open-Source English model (F1: 0.573, Balanced Accuracy: 0.676 vs. F1: 0.587, Balanced Accuracy: 0.627). The GPT English model shows moderate performance but still falls short of the Swedish-specific models.
Because of its higher F1 score and Balanced Accuracy, we use the Swedish GPT sentiment scoring as the baseline for the remainder of this paper. The relatively high accuracies obtained here confirm \cite{holmstrom2023bridging}'s findings, i.e.  LLMs can be used for ``smaller'' languages, specifically Swedish, without extensive fine-tuning.

\begin{table}[H]
\centering
\caption{Model Performance Metrics: F1 Score and Balanced Accuracy}
\begin{tabular}[t]{lcc}
\toprule
Prediction Method & F1 Score & Balanced Accuracy\\
\midrule
Open-Source English & 0.587 & 0.627\\
Open-Source Swedish & 0.573 & 0.676\\
GPT English & 0.575 & 0.603\\
GPT Swedish & 0.696 & 0.696\\
\bottomrule
\end{tabular}
\label{Tab: Sentiment}
\end{table}

\subsection{Estimating Politician Ideology}
\label{sc:polid}

To systematically analyse the relationship between ideology and affective polarization, we need reliable and comparable measures of political ideology. To achieve this, we use three separate approaches to quantify the ideologies of the political figures in our sample: i) the DW-NOMINATE framework applied to the Swedish Parliament, ii) the Chapel Hill Expert Survey (CHES) ideological ratings of political parties and, iii) the LLM assessment of party ideologies. All computed values range from -1 (left) to 1 (right), with 0 representing the political centre \footnote{DW-NOMINATE is a notable exception, as it is a standalone estimation method rather than a direct measure of ideology. Our analysis finds that a negative DW-NOMINATE value positively correlates with a right-leaning ideology, whereas a negative value correlates with a left-leaning ideology.}.

\begin{enumerate}

\item \textbf{DW-NOMINATE.} The DW-NOMINATE framework is a statistical method used to estimate the ideological positions of legislators based on their roll-call voting behaviour. Its dimensions represent i) the liberal-conservative divide and ii) societal issues such as civil rights \citep{poole2011ideology}.  \\
Though applications of the DW-NOMINATE exist in Europe \citep{hix2005power}, they are typically less prevalent. 
This can mainly be explained by the presence of the multi-party systems in the EU and its resulting political coalitions. To a lesser extent, the ease of access to European roll-call vote data dissuades its applications. \\
We compute the DW-NOMINATE score for the Swedish parliament by downloading the history of the roll call votes of the last 10 years (2014 - 2024). We subsequently apply the DW-NOMINATE procedure \citep{poole2011ideology} on the collected data. Of the $78$ politicians in the sample, $17$ were not represented in parliament within the past 10 years. We solely use the first dimension of DW-NOMINATE as it typically captures the left-right ideology spectrum.

\item \textbf{Chapel Hill Expert Survey.} The CHES estimates party positioning on European integration, ideology and policy issues for national parties in most European countries \citep{jolly2022chapel}. Though the data contains a lot of information on parties' views on the European Union, immigration and other prominent political topics, we chose to solely include the general stance of the party from extreme left to extreme right. \\
These ratings allow us to establish the political stance through an informed expert opinion. Nevertheless, they come with two major downsides. First, the last survey was carried out in 2019, in the meantime some political parties might have shifted their ideologies. Second, the expert ratings only provide party-wide estimates with additional information on the political figures in the party.

\item  \textbf{LLM Estimation.} We also use LLM (Large Language Models) to assess the ideologies of the political parties and politicians in our sample. Specifically, we query \texttt{GPT-4o}\footnote{OpenAI uses its latest model when no specific checkpoint is specified. Specifically, we used \texttt{gpt-4o-2024-08-06}. OpenAI will likely discontinue this version in favour of newer models shortly after the publication of this paper.} with the following query: \textit{``As an AI model with extensive knowledge in Swedish politics, rate this political party from -1 (extreme left) to 1 (extreme right) with 2 decimal numbers. Answer only with a number in the aforementioned range''}. An analogue query was used to get the rating of the politicians instead of the political party \cite{tornberg2024large}. \\
LLMs enable us to rate the political extremity of many politicians and their parties easily. Beyond their convenience, modern LLMs have been trained on vast and diverse text corpora, giving them a nuanced understanding of language, context, and ideological subtext. In principle, they should capture the average feeling towards the Swedish political parties and politicians, thereby including the opinion of the Swedish citizens in our extremity metrics. Something traditional metrics such as DW-NOMINATE fully exclude. 

\end{enumerate}

It is worth noting that \cite{yu2024partisanship} also use the spatial following model presented by \cite{barbera2015birds}. Given the changes made to the Twitter API in 2023, we cannot replicate this. Other models also exist within the literature, namely \cite{mosleh2022measuring} use elite account following to quantify the partisanship of an account. Unfortunately, such models are often grounded in U.S. politics with limited applicability outside of their native context.

\section{Results and Analysis} \label{sc:results}

Our initial expectation was that political elites would engage more frequently in negative partisanship (expressing negative sentiments toward out-group politicians) than positive partisanship (expressing positive sentiments toward in-group politicians). To test this, we conducted a Bayesian Poisson Rate Comparison Test \citep{burkner2021bayesian}, employing Jeffreys priors (uninformative) for the Poisson rate parameters. The test computes the likelihood of a politician to post a positive tweet given it is in-group and the likelihood to post a negative sentiment tweet given it is out-group.
We computed the statistics under two definitions of group membership. First, by defining a group as a political party where all out-of-party tweets are out-of-group tweets. Second, by defining a group as a political bloc (left-wing or right-wing). As outlined in section \ref{sc:TweetSent} and in Table \ref{Tab: Sentiment}, the analysis utilises OpenAI Swedish sentiment given its superior F1-score and Balanced Accuracy.\footnote{We note that all sentiment models yield similar findings. The computations using the other metrics can be found on the GitHub page.}

The results indicate that the balance between negative and positive partisanship 
$\left[ \frac{\lambda_{\text{positive}}}{\lambda_{\text{negative}}}\right]$,
varies depending on the definition of the in-group. When the in-group is defined as the broader in-bloc, the difference between positive and negative partisanship is minimal, with positive partisanship being slightly more prevalent ($RateRatio = 0.94 [.90,.99]$). In contrast, when the in-group is defined more narrowly as the in-party, negative partisanship dominates ($RateRatio = 1.51 [1.45,1.58]$).

\begin{table}[H]
\centering
\caption{Likelihood of Negative and Positive Partisan Tweets (H1).}
\begin{tabular}{lrrrrr|rrrrr}
\toprule
\multicolumn{1}{c}{} & \multicolumn{5}{c}{Bloc} & \multicolumn{5}{c}{Party} \\ 
\cmidrule(lr){2-6} \cmidrule(lr){7-11}
Parameter & Mean & SD & 2.5\% & 50\% & 97.5\% & Mean & SD & 2.5\% & 50\% & 97.5\% \\ 
\midrule
NP likelihood      & 0.49 & 0.01 & 0.47 & 0.49 & 0.50  & 0.60 & 0.01 & 0.59 & 0.60 & 0.62\\ 
PP likelihood      & 0.52 & 0.01 & 0.50 & 0.51 & 0.53 & 0.40 & 0.01 & 0.38 & 0.40 & 0.41  \\ 
RateRatio          & 0.94 & 0.02 & 0.90 & 0.94 & 0.99 & 1.51 & 0.03 & 1.45 & 1.51 & 1.58 \\ 
RateDifference     & -0.03 & 0.01 & -0.05 & -0.03 & -0.01 & 0.20 & 0.01 & 0.18 & 0.20 & 0.23 \\
\bottomrule
\end{tabular}
\vspace{2pt}
\caption*{\textit{Note: Summary of In-Bloc and In-Party Poisson parameters. For all parameters $N_{eff} \geq 400$ and $\hat{R} \leq 1.01$, $N = 7594$. The Model ran with 6 chains of 2000 iterations (of which 1000 warm-ups).}}
\label{tab:poisson}
\end{table}

\begin{figure}
    \centering
    \caption{Negative and Positive Tweets for the In- and Out-Groups}
    \begin{subfigure}{0.45\textwidth}
        \centering
        \includegraphics[width=\linewidth]{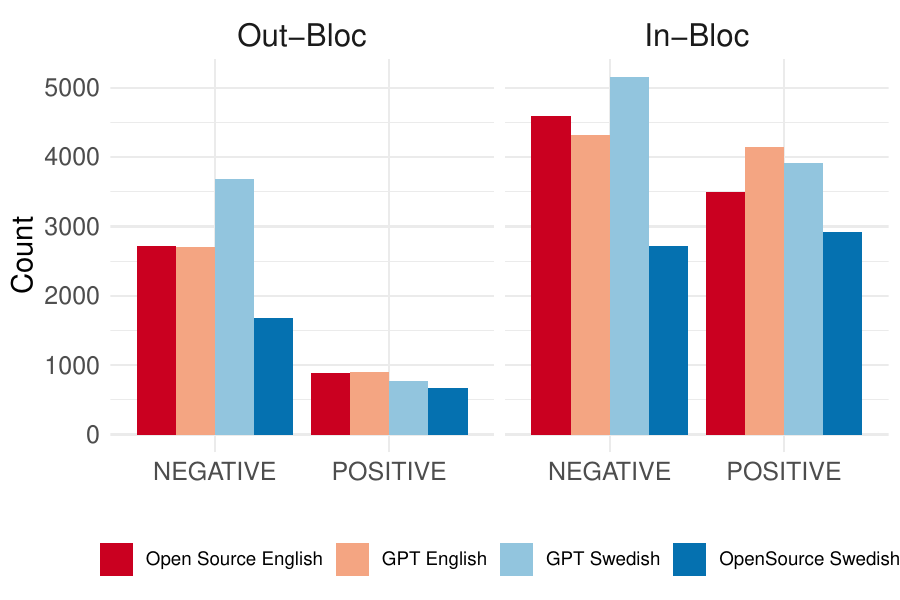}
        \caption{Bloc as In-Group}
        \label{fig:bloc_rate}
    \end{subfigure}
    \hfill
    \begin{subfigure}{0.45\textwidth}
        \centering
        \includegraphics[width=\linewidth]{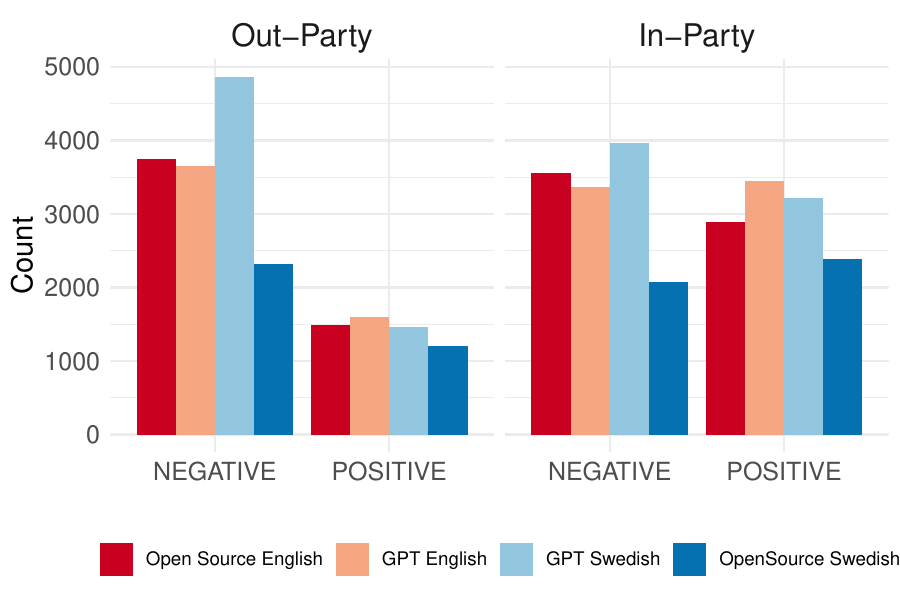}
        \caption{Party as In-Group}
        \label{fig:party_rate}
    \end{subfigure}
    
    \caption*{\textit{Note: The bar charts display the total Negative and Positive tweets posted by the politicians. It excludes the Neutral sentiment. Plot ``a'' and ``b'' respectively define the Bloc and the Party as in the in-group.The colours are indicative of the sentiment classifiers.}}
    \label{fig:posneg_tweets}
\end{figure}

\begin{figure}
\centering
\caption{Tweet Weighted Network}
\includegraphics[width=0.5\textwidth]{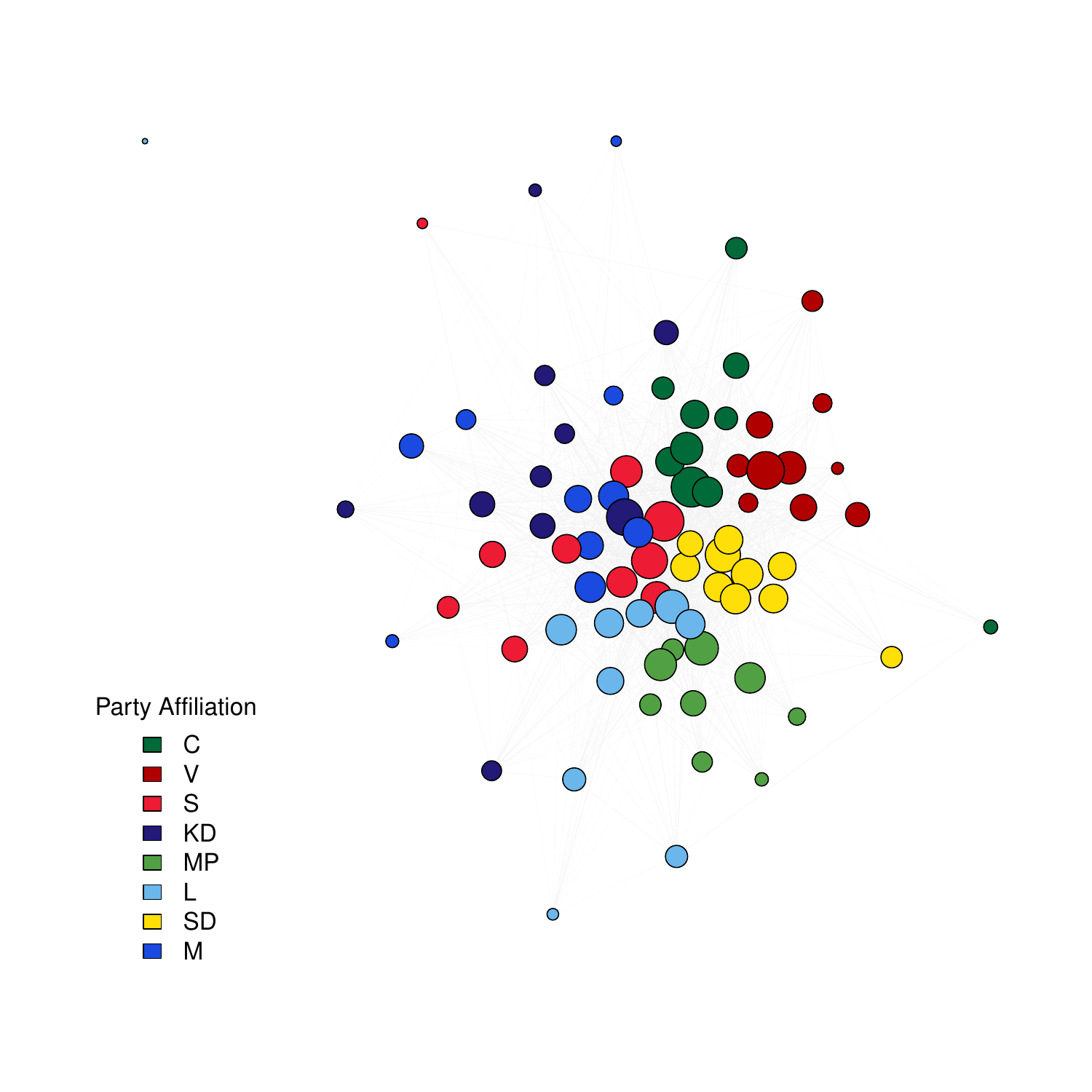}
\caption*{\textit{Note: Weighted Network of the Tweets posted by the Swedish Politicians. A positive mention brings two nodes (politicians) closer; a negative mention pushes them further apart.}}
\label{fig:network}
\end{figure}

Figure \ref{fig:posneg_tweets} illustrates this finding: when defining the group as the in-bloc, the rates of negative and positive partisanship are nearly equivalent. In contrast, under the in-party definition, negative partisanship is more prevalent. Moreover, as shown in Figure~\ref{fig:network}, the network structure of Swedish tweets is more strongly influenced by party affiliations than by bloc alignments. In this network, positive mentions draw politicians closer together, whereas negative mentions push them further apart. i.e. a homogenous party cluster implies a high free of affective polarisation. The pattern at hand indicates that politicians predominantly support members of their own party, with the underlying bloc structure being largely absent. It is also worth noting that the degree of homogeneity varies across parties. Overall, our findings provide clear support for our first hypothesis in the in-party setting, but not in the in-bloc setting.

Our second hypothesis suggested that negative partisanship is more likely to gather reactions than its positive counterpart. To test this relationship we run a hierarchical Bayesian mixed-effects model described by equations \ref{eq:h2start} to \ref{eq:h2end}. The models estimate the number of logged likes or retweets a tweet gets based on its sentiment and the political leanings of its author and the tweet's targeted politician (i.e. the politician mentioned in the tweet).  We control for the random effects of the author to account for individual-level differences in engagement, and the hour of posting as tweets posted during peak hours are more likely to gather likes and retweets. The model features 8 chains of 2000 iterations (of them 1000 warm-ups). For all parameters $\hat{R} \leq 1.01$ and $N_{eff} \geq 400$. These crude metrics indicate that the model has converged adequately, with low between-chain variance and a sufficiently large effective sample size, suggesting reliable posterior estimates.

Unlike \cite{yu2024partisanship}, who included only negative and positive partisanship tweets, we include all tweets posted by politicians, irrespective of their target and sentiment. This decision is motivated by the observation that the bulk of a politician's tweets tend to be neutral rather than exclusively centred on praise or criticism. 

All models consistently support Hypothesis 2, i.e., negative partisanship generates more likes and retweets than positive partisanship.
Table \ref{tab:H2} displays the results of all regressions using either retweets or likes and either political party or political bloc as the in-group definition. Since Table \ref{tab:H2} reports the estimated logged values of engagement (likes and retweets), we also provide the exponentiated values to facilitate interpretation in terms of actual engagement counts.
Using political parties as the in-groups, a positive partisan will get only a relatively low amount of reactions, with $2.86$ likes $[1.66,4.99]$ and $8.96$ retweets $[5.26,14.87]$. Ceteris paribus, a negative partisan tweet will gather $9.11$ likes $[5.30,15.88]$ and $15.10$ retweets $[8.86,24.96]$. The same pattern holds in the bloc as in-group setting\footnote{In the bloc setting: A positive partisan tweet will gather $2.09$ likes $[1.63,2.75]$ and $6.27$ retweets $[4.76,8.31]$. For a negative partisan tweet, $7.87$ likes $[6.10,10.35]$ and $12.57$ retweets $[9.58,16.70]$}.  It is noteworthy that all politicians, regardless of their position on the political spectrum, experience increased reactions from negative partisanship on average. However, the precise increase in likes and retweets varies by individual and party. For illustration, we compute the estimates for the \textit{Swedish Social Democratic Party (S)}, which held the most seats during the period of the sample.

To further illustrate the impact of negative partisanship on social media engagement, Figure \ref{fig:h2plot} displays the posterior distributions of the reactions ratio $\left[\frac{NP Reactions}{PP Reactions}\right]$. It illustrates how much more likes (or retweets) a tweet from a negative partisan receives relative to a tweet from a positive partisan in the same in-group setting, e.g. in the party-as-in-group setting, a negative partisan tweet would on average, gather $3.18$ more likes $[2.95, 3.44]$ and $1.69$ more retweets $[1.57,1.81]$ than its positive counterpart.

\begin{align}
\label{eq:h2start}
Reactions_i = &  \alpha + \beta_1 PolAffiliation_{i} + \beta_2 InGroup_{i} +\beta_3 PP_{i} + \beta_4 NP_{i} + \\
& \gamma^{author}_{a[i]} + \gamma^{time}_{t[i]} \notag  \\ 
\alpha \sim & \mathcal{N}(0,1); \\
\bm{\beta} \sim &  \mathcal{N}(0,1); \\
\gamma^{u} \sim & \mathcal{N}(0, \sigma^{u}), \hspace{10pt}\ \sigma^{u} \sim  \mathcal{N}^{+}(0,1), \hspace{10pt}\forall \hspace{2pt} u \in \{\text{author, time}\};
\label{eq:h2end}
\end{align}

\begin{table}[p]
\caption{Positive and Negative Partisanship on Likes and Retweets (H2).}
\caption*{\textit{$Note$: The table reports mean, standard deviation, and 95\% credible intervals (2.5\% and 97.5\%) for the model parameters. The author and time covariates are not shown in the table but are available in the online appendix. The reference sentiment is ``Neutral''.}} 

\begin{subtable}[t]{\textwidth}
\label{tab:H2bloc}
\centering
\caption{Political Bloc as Group}
\begin{tabular}{l rrrr|rrrr}
  \hline
  Parameter &\multicolumn{4}{c}{\textbf{log(Likes)}}&\multicolumn{4}{c}{\textbf{log(Retweets)}}\\
  \cmidrule(lr){2-5}  \cmidrule(lr){6-9}
   & 2.5\% & 50\% & 97.5\% & $P\geq0$ & 2.5\% & 50\% & 97.5\%  & $P\geq0$\\
  \hline
Intercept & 0.09 & 0.34 & 0.62 & 1.00 & 1.80 & 2.07 & 2.35 & 1.00 \\ 
    \textbf{OutBloc} & 0.71 & 0.78 & 0.85 & 1.00 & -0.06 & 0.00 & 0.06 & 0.54 \\ 
    \textbf{OutBloc:Negative} & 0.63 & 0.73 & 0.83 & 1.00 & -0.24 & -0.15 & -0.06 & 0.00 \\ 
    \textbf{InBloc:Positive} & -0.01 & 0.14 & 0.29 & 0.97 & 0.29 & 0.42 & 0.55 & 1.00 \\ 
    RightBloc & 0.23 & 0.61 & 0.95 & 1.00 & -0.02 & 0.34 & 0.70 & 0.97 \\ 
    \textbf{NegativeSentiment} & 0.15 & 0.21 & 0.27 & 1.00 & 0.56 & 0.61 & 0.66 & 1.00 \\ 
    \textbf{PositiveSentiment} & 0.12 & 0.25 & 0.38 & 1.00 & -0.78 & -0.66 & -0.54 & 0.00 \\ 
  \hline
  
\end{tabular}

\end{subtable}

\vspace{1em}

\begin{subtable}[t]{\textwidth}
\label{tab:H2party}
\centering
\caption{Political Party as Group}
\begin{tabular}{l rrrr|rrrr}
  \hline
  Parameter &\multicolumn{4}{c}{\textbf{log(Likes)}}&\multicolumn{4}{c}{\textbf{log(Retweets)}}\\
  \cmidrule(lr){2-5}  \cmidrule(lr){6-9}
   & 2.5\% & 50\% & 97.5\%  &$P\geq0$& 2.5\% & 50\% & 97.5\%  &$P\geq0$\\
  \hline
 Intercept & 0.14 & 0.68 & 1.23 & 0.99 & 1.83 & 2.35 & 2.85 & 1.00 \\ 
    \textbf{OutParty} & 0.66 & 0.72 & 0.79 & 1.00 & -0.17 & -0.11 & -0.06 & 0.00 \\ 
    \textbf{OutParty:Negative} & 0.54 & 0.63 & 0.73 & 1.00 & -0.26 & -0.18 & -0.09 & 0.00 \\ 
    \textbf{InParty:Positive} & -0.07 & 0.05 & 0.17 & 0.79 & 0.38 & 0.49 & 0.59 & 1.00 \\ 
    Party:C & -1.10 & -0.37 & 0.36 & 0.15 & -1.39 & -0.69 & 0.03 & 0.03 \\ 
    Party:KD & -0.71 & 0.01 & 0.75 & 0.51 & -0.39 & 0.30 & 1.01 & 0.81 \\ 
    Party:L & -0.62 & 0.14 & 0.91 & 0.64 & -1.57 & -0.84 & -0.10 & 0.01 \\ 
    Party:M & -0.43 & 0.32 & 1.04 & 0.81 & -0.30 & 0.37 & 1.07 & 0.85 \\ 
    Party:MP & -1.36 & -0.63 & 0.13 & 0.05 & -0.97 & -0.27 & 0.44 & 0.23 \\ 
    Party:SD & -0.33 & 0.39 & 1.13 & 0.86 & -0.17 & 0.52 & 1.25 & 0.92 \\ 
    Party:V & -1.36 & -0.63 & 0.11 & 0.05 & -0.63 & 0.05 & 0.74 & 0.55 \\ 
    \textbf{NegativeSentiment} & 0.11 & 0.18 & 0.25 & 1.00 & 0.59 & 0.65 & 0.71 & 1.00 \\ 
    \textbf{PositiveSentiment} & 0.23 & 0.32 & 0.42 & 1.00 & -0.73 & -0.65 & -0.56 & 0.00 \\ 

   \hline
\end{tabular}
\caption*{\textit{Note: The Swedish Social Democratic Party (S) was chosen as the reference category for Party as it has the most MPs in the Swedish parliament.}}
\end{subtable}
\label{tab:H2}
\end{table}

\begin{figure}[h]
    \centering
    \caption{Posterior distribution of the ratio of negative to positive partisanship reactions}
    \includegraphics[width=0.95\linewidth]{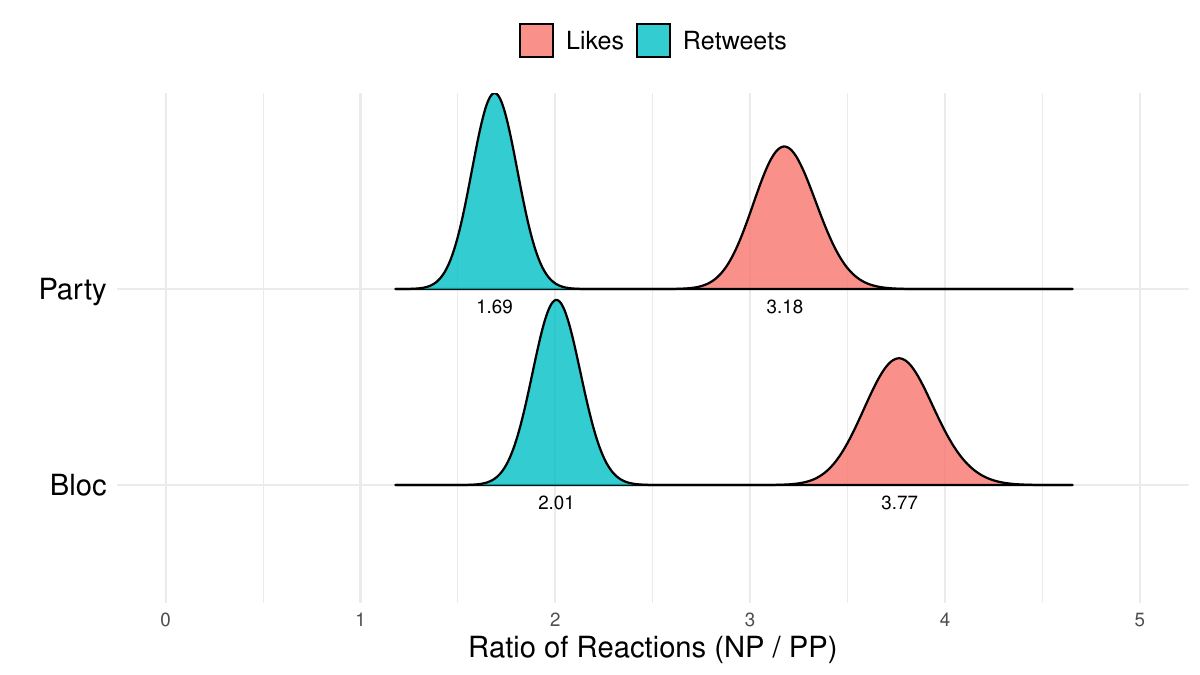}
    \caption*{\textit{Note: The plot displays the posterior distributions of the ratio of estimated engagement (likes and retweets) between negative and positive partisanship. The density ridges illustrate the distribution of these ratios across different reaction types (likes and retweets) and categories (party and bloc). The ratio is computed by dividing the estimated reactions for negative partisanship by the estimated engagement for positive partisanship. These estimates are derived from the regression model, where engagement is predicted as a function of sentiment, partisanship, and their interaction. Note that the sum of the coefficients is exponentiated as the model predicts logged reactions.}}

    \label{fig:h2plot}
\end{figure}

Hypothesis 3 suggests that more extreme politicians are likelier to engage in negative partisanship than moderates. To test this, we use a hierarchical Bayesian logistic regression model to predict the likelihood that a tweet contains negative partisanship. The regression uses the estimated political ideologies detailed in section \ref{sc:polid} as independent variables. 
Note that the party or bloc affiliations are excluded from this regression as they are perfectly co-linear with the estimated party ideology ratings. The model is formally defined in equations \ref{eq:h3start} to \ref{eq:h3end}. 
To examine the impact of political ideology on the likelihood of negative partisanship, we include both the \textit{IdeologyMetric} term and its square. The non-squared term captures the general relationship between left-right ideological positions and negative partisanship, while the squared term accounts for the increased likelihood of negative partisanship among individuals with more extreme ideological positions, whether on the left or right\footnote{As detailed in section \ref{sc:method}, the ideology goes from -1 (left) to 1 (right)}.
Additionally, we control for the random effects of post timing and author\footnote{The significant positive effects of Likes and Retweets across models further support H2.}.
The model runs 8 chains of 2,000 iterations (including 1,000 warm-ups). For all parameters, $\hat{R} \leq 1.02$ and $N_{eff} \geq 400$, ensuring convergence. The results are presented in Table \ref{tab:H3}.

We find partial support for Hypothesis 3.  
Using the DW-NOMINATE metric, we find no evidence that a politician’s ideological extremity affects the likelihood of posting a negative partisan tweet. This is likely due to DW-NOMINATE’s limitations in capturing ideological dynamics in Sweden\footnote{DW-NOMINATE's non-squared term suggests that right-leaning politicians are more likely to spread fake news. However, we do not interpret this result for two reasons: (1) it does not capture the effect of ideological extremity on negative partisanship, and (2) DW-NOMINATE does not strictly measure ideology on a conventional left-right spectrum. As detailed in section \ref{sc:polid}, a negative value for DW-NOMINATE correlates with a right-leaning ideology}.
Similarly, there is only weak support for the relationship between negative partisanship and political extremity when defining the in-group at the bloc level.

In contrast, we find consistent support for H3 when using a party-based definition of the in-group. The \textit{IdeologyMetric} term and its square suggest that more moderate politicians are less likely to post negative partisanship, while both extreme left and right politicians are more likely to do so. 
As with H1, this finding suggests that party identity plays a more significant role in shaping partisan behaviour than broader ideological positioning. Still similarly to H1, this effect is contingent on how the in-group is defined, achieving significance only when the in-group is operationalized at the party level.
Using the CHES-based ideology we find that the \textit{Center Party (C)} was the least likely to engage in negative partisanship with an estimated likelihood to engage in negative partisanship of $0.10$ $[0.08,0.13]$ whilst the \textit{Sweden Democrats (SD)}, a right-wing populist party, were the most likely $0.19$ $[0.15,0.26]$. That is, SD, the most negative-partisan party, was $1.86$ more likely to engage in negative-partisanship than C, the least negative-partisan party $[1.37,2.62]$\footnote{Across all ideology metrics, C and SD were consistently among the least and most likely to engage in negative partisanship.}.

\begin{align}
\label{eq:h3start}
\bm{\nu}_{i} \sim & \mbox{Bernouilli}(\pi_{i}) \\
\mbox{logit}(\pi_i) = & \alpha + \beta_1 IdeologyMetric_{i} + \beta_2 IdeologyMetric_{i}^{2} + \\
& \beta_3 Likes_i +\beta_4 Retweets_{i} + \gamma^{author}_{a[i]} + \gamma^{time}_{t[i]} \notag \\ 
\alpha \sim & \mathcal{N}(0,1) \\
\bm{\beta} \sim &  \mathcal{N}(0,1) \\
\gamma^{u} \sim & \mathcal{N}(0, \sigma^{u}), \hspace{10pt}\ \sigma^{u} \sim  \mathcal{N}^{+}(0,1), \hspace{10pt}\forall \hspace{2pt} u \in \{\text{author, time}\}
\label{eq:h3end}
\end{align}

\begin{table}[p]
    \centering
    \footnotesize
    \caption{Likelihood of Posting Negative Partisan Tweets (H3)}
    \begin{center} 
    \begin{tabular}{cl SSSS|SSSS} 
        \toprule
        \textbf{\shortstack{Ideology \\ Metric}} & & \multicolumn{4}{c}{\textbf{Party}} & \multicolumn{4}{c}{\textbf{Bloc}} \\  
        \cmidrule(lr){3-6} \cmidrule(lr){7-10}
        \multicolumn{2}{c}{} & {2.5\%} & {50\%} & {97.5\%} & {$P\geq0$} & {2.5\%} & {50\%} & {97.5\%} &  {$P\geq0$} \\  
        \midrule

        \multirow{5}{*}{\parbox{2cm}{\centering \text{CHES}}}
        & Intercept    & -2.36 & -2.00 & -1.67 & 0.00 & -2.62 & -2.29 & -1.95 & 0.00 \\  
        & Likes        & 0.56 & 0.59 & 0.63 & 1.00 & 0.61 & 0.65 & 0.69 & 1.00 \\  
        & Retweets     & 0.22 & 0.25 & 0.29 & 1.00 & 0.21 & 0.25 & 0.29 & 1.00 \\  
        & \textbf{IdeologyMetric} & -0.61 & -0.24 & 0.12 & 0.09 & -0.30 & 0.07 & 0.43 & 0.65 \\  
        & \textbf{IdeologyMetric\textsuperscript{2}} & -0.03 & 1.04 & 2.23 & 0.97 & -0.64 & 0.50 & 1.60 & 0.79 \\  
        \midrule

        \multirow{5}{*}{\parbox{2cm}{\centering \text{GPT} \\ \text{Politician}}}
        & Intercept    & -2.33 & -1.98 & -1.63 & 0.00 & -2.70 & -2.37 & -2.04 & 0.00 \\  
        & Likes        & 0.56 & 0.59 & 0.62 & 1.00 & 0.61 & 0.65 & 0.68 & 1.00 \\  
        & Retweets     & 0.21 & 0.25 & 0.29 & 1.00 & 0.21 & 0.25 & 0.29 & 1.00 \\  
        & \textbf{IdeologyMetric} & -0.38 & -0.07 & 0.23 & 0.33 & -0.25 & 0.05 & 0.35 & 0.64 \\  
        & \textbf{IdeologyMetric\textsuperscript{2}} & -0.20 & 0.60 & 1.44 & 0.94 & -0.17 & 0.62 & 1.40 & 0.94 \\  
        \midrule

        \multirow{5}{*}{\parbox{2.2cm}{\centering \text{GPT} \\ \text{Party}}}
        & Intercept    & -2.37 & -2.02 & -1.69 & 0.00 & -2.67 & -2.34 & -1.99 & 0.00 \\  
        & Likes        & 0.56 & 0.59 & 0.62 & 1.00 & 0.61 & 0.65 & 0.68 & 1.00 \\  
        & Retweets     & 0.21 & 0.25 & 0.29 & 1.00 & 0.21 & 0.25 & 0.29 & 1.00 \\  
        & \textbf{IdeologyMetric} & -0.24 & 0.05 & 0.34 & 0.64 & -0.10 & 0.21 & 0.50 & 0.92 \\  
        & \textbf{IdeologyMetric\textsuperscript{2}} & -0.08 & 0.65 & 1.37 & 0.96 & -0.21 & 0.49 & 1.21 & 0.91 \\  
        \midrule

        \multirow{5}{*}{\parbox{2cm}{\centering \text{DW-} \\ \text{NOMINATE}}}
        & Intercept    & -1.98 & -1.68 & -1.42 & 0.00 & -2.30 & -1.99 & -1.71 & 0.00 \\  
        & Likes        & 0.58 & 0.62 & 0.65 & 1.00 & 0.64 & 0.68 & 0.72 & 1.00 \\  
        & Retweets     & 0.22 & 0.26 & 0.30 & 1.00 & 0.20 & 0.25 & 0.30 & 1.00 \\  
        & \textbf{IdeologyMetric} & -0.76 & -0.36 & 0.03 & 0.04 & -0.79 & -0.38 & 0.03 & 0.04 \\  
        & \textbf{IdeologyMetric\textsuperscript{2}} & -0.86 & -0.10 & 0.67 & 0.39 & -1.25 & -0.39 & 0.45 & 0.19 \\  
        \bottomrule
    \end{tabular}
    \end{center} 
    \caption*{\textit{Note: The table reports 95\% credible intervals (2.5\% and 97.5\%) for the model parameters. Results are shown separately for Party and Bloc in-group definitions across four ideology metrics. The dependent variable is the likelihood of posting a negative partisan tweet. $P\geq0$ represents the probability that the coefficient is greater than or equal to zero. The median estimate is the substitute for the point estimate in a frequentist setting.}}
    \label{tab:H3}
\end{table}

\begin{figure}[h]
    \centering
    \caption{Estimated probability of posting a negative partisan tweet by party and ideological rating}  
    \includegraphics[width=0.95\linewidth]{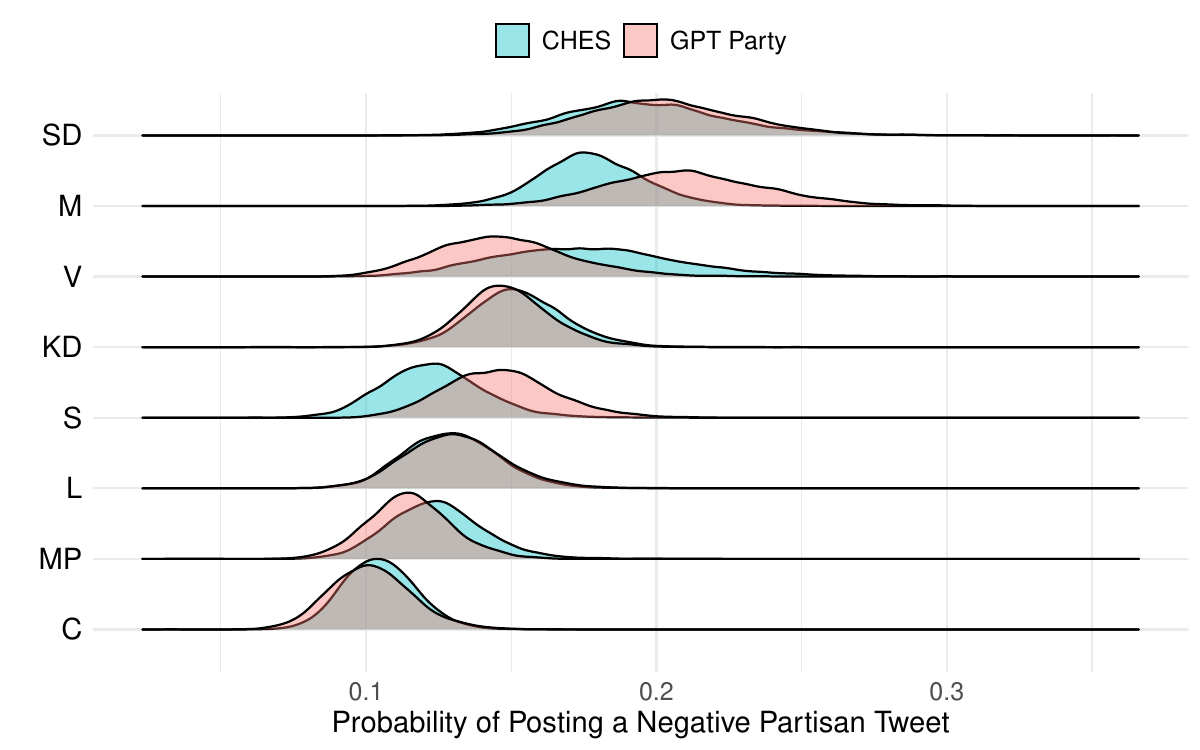}  
    \caption*{\textit{Note: The figure presents posterior distributions of the probability of posting a negative partisan tweet by party. The x-axis represents the estimated probability, while the y-axis orders parties by their median probability. The probabilities were calculated using the posterior distribution of the regression coefficients for both CHES and GPTparty ratings.}}  
    \label{fig:h2plot}  
\end{figure}

\section{Discussion} \label{sc:discussion}

This study had two main goals. First, it examined how and why political elites engage in positive and negative partisanship in a multi-party system such as Sweden. By exploring these dynamics, we aim to understand how partisanship evolves and the factors that drive people's political behaviour. Second, it explored whether new LLM technologies can help analyse lesser-studied, non-anglophone countries like Sweden, where datasets like VoteView or institutions such as the ANES aren't as readily available. 

\subsection{Partisanship Online}

This study replicates the methods and objectives of \citet{yu2024partisanship}, though we differ significantly from it in its operationalisation, setting and findings. 

Specifically, the balance between negative and positive partisanship depends on how political in-groups are defined. When the in-group is defined as the broader left- or right-bloc, negative and positive partisanship occur at similar rates, with positive partisanship being slightly more prevalent. This pattern is reminiscent of the United States, where positive partisanship remains the dominant narrative despite the rise of negative partisanship \citep{lee2022negative, yu2024partisanship, iyengar2018strengthening}. However, when the in-group is narrowed to the party level, negative partisanship becomes significantly more dominant. This suggests that while ideological blocs may encourage some degree of unity, party membership generates stronger patterns of opposition and criticism in online discourse.

Negative partisanship’s increased visibility is likely a key reason why politicians engage in it. Since tweets expressing negative partisanship receive more engagement than positive ones, politicians may have an incentive to adopt a more critical tone to maximize their reach \citep{schone2021negativity}. Regardless of the in-group definition used, whether in-party or in-bloc, we observe substantial increases in likes and retweets when Swedish elite politicians use negative partisanship. On average, a negative partisan tweet will gather $3.18$ and $3.77$ more likes in the in-party and in-bloc settings, respectively.

As with Hypothesis 1, our results show that the effect of ideological extremity on negative partisanship depends on how the in-group is defined. When the in-group is framed as a broader ideological bloc (left or right), we find no conclusive evidence that ideological extremity—whether left or right—influences the likelihood of posting a negative partisan tweet. This suggests that broad ideological affiliations may foster some unity for extreme politicians, reducing the need for overt negative partisanship. However, when defining the in-group at the party level, we find that ideologically extreme politicians are significantly more likely to engage in negative partisanship. The \textit{Sverigedemokraterna (SD)}, an extreme-right party, being $1.86$ more likely to engage in Negative than the \textit{Center Party (C)}, a Centrist party This finding aligns with \citet{yu2024partisanship}, who show that ideological extremity amplifies partisan hostility but only when the in-group is defined at the party level. Similarly, it suggests that, as with voters, individuals with stronger partisan identities and more extreme political attitudes tend to exhibit greater affective polarization \cite{reiljan2021ideological}.

The underlying thesis of this paper has been that affective polarization manifests itself differently in multiparty systems compared to two-party systems. Our results confirm that the definition of the in-group is crucial in shaping patterns of affective polarization, with statistical significance contingent on how these political identities are framed. Specifically, we find that in multiparty systems, party-based identities foster more intense divisions and interparty competition, amplifying the role of affective polarization—particularly negative partisanship. 
This aligns with research suggesting that partisan identity often functions as a social identity, reinforcing group cohesion and intergroup animosity \citep{mason2018uncivil, iyengar2019origins}. In contrast, broader ideological groupings, such as left- or right-blocs, may not provoke the same level of partisan hostility, as they lack the distinctiveness and exclusivity of party-based identities.

This pattern can be partially explained by the dynamics of smaller, more cohesive groups, which tend to exhibit stronger in-group solidarity and a heightened perception of threat from out-groups\citep{huddy2015expressive}. Research on affective polarization in multiparty contexts suggests that when political identities are more narrowly defined—such as by party rather than ideological bloc—affective polarization becomes more pronounced \citep{widmann2021emotional, westwood2018tie}. This effect is particularly strong among extremists, whose ideological positions diverge sharply from those of moderates, reinforcing negative partisanship as a mobilizing force \citep{tajfel1979integrative}.

\subsection{LLMs for Data Labelling}

A significant challenge in researching negative partisanship, particularly in multiparty systems like Sweden, is the lack of readily available data tailored to such contexts. Many established resources, such as the VoteView database and the American National Election Studies, lack European equivalents \cite{lewis2019voteview}. In an attempt to adapt existing tools, we constructed our own DW-NOMINATE metric using the original framework applied to Swedish roll-call data. However, we found that DW-NOMINATE did not perform well in this context. This is likely due to the Swedish political environment, which deviates from the model’s core assumptions. Specifically, DW-NOMINATE is built on the premise of a two-dimensional political space with clearly separated political blocs. An assumption that does not align with the more fluid, multiparty landscape of Sweden and many other European countries. It further highlights the need for EU-specific metrics \citep{wagner2024affective}.

In contrast, recent advances in LLMs have made it easier to classify political discourse on a large scale, yielding results consistent with expert-grounded assessments like CHES. By leveraging GPT for political classification, we also benefit from the flexibility and scalability of LLMs, bypassing the need for extended roll-call data or expert opinions. Our work contributes to the growing literature on the use of LLMs in political analysis, building on recent findings that LLMs like GPT-4 can accurately classify political affiliations in social media content \citep{tornberg2024large}.

Beyond political classification, we find that LLMs can also accurately classify the sentiment of political tweets in a Swedish context. Specifically, \texttt{GPT-4o} demonstrates a high level of precision, achieving an F1 score of 0.69 across three sentiment categories when analyzing Swedish tweets without any model fine-tuning. Notably, LLMs tend to perform better in their native language, further enhancing accuracy. While open-source alternatives offer greater transparency compared to OpenAI's \texttt{GPT-4o}, we found their accuracy to be generally lower. However, the difference in performance between licensed and open-source models was less significant than the impact of the tweet's language (native or translated English).

\newpage

\bibliography{references.bib}

\end{document}